\documentclass{emulateapj}

\usepackage{wasysym}

\newcommand{\figw}{\columnwidth}
\def\kms{km.s$^{-1}$}         
\def\ms{\hbox{m.s$^{-1}$}}         
\def\kms{\hbox{km.s$^{-1}$}}       
\def\vsini{\hbox{$\upsilon \sin i_{\star}$}}      
\def\Msun{\hbox{$\mathrm{M}_{\astrosun}$}}             
\def\Rsun{\hbox{$\mathrm{R}_{\astrosun}$}}
\def\Mjup{\hbox{$\mathrm{M}_{\jupiter}$}}
\def\Rjup{\hbox{$\mathrm{R}_{\jupiter}$}}

\def\degr{\hbox{$^\circ$}}
\def\teff{T$_{\rm eff}$}
\def\logg{log~{\it g}}
\def\met{[Fe/H]}

\def\vmacro{$\upsilon_\mathrm{macro}$}

\shorttitle{EPIC211089792~b}
\shortauthors{Santerne et al.}

\begin{document}

\title{EPIC211089792~b: an aligned and inflated hot jupiter in a young visual binary}

\author{A.~Santerne\altaffilmark{1,2}}
\email{alexandre.santerne@astro.up.pt}
\author{G.~H\'ebrard\altaffilmark{3,4}}
\author{J.~Lillo-Box\altaffilmark{5,6}}
\author{D.~J.~Armstrong\altaffilmark{7,8}}
\author{S.~C.~C.~Barros\altaffilmark{1}}
\author{O.~Demangeon\altaffilmark{2}}
\author{D.~Barrado\altaffilmark{5}}
\author{A.~Debackere\altaffilmark{9}}
\author{M.~Deleuil\altaffilmark{2}}
\author{E.~Delgado~Mena\altaffilmark{1}}
\author{M.~Montalto\altaffilmark{1}}
\author{D.~Pollacco\altaffilmark{7}}
\author{H.~P.~Osborn\altaffilmark{7}}
\author{S.~G.~Sousa\altaffilmark{1}}

\author{L.~Abe\altaffilmark{10,9}}
\author{V.~Adibekyan\altaffilmark{1}}
\author{J.-M.~Almenara\altaffilmark{11,12}}
\author{P.~Andr\'e\altaffilmark{13,9}}
\author{G.~Arlic\altaffilmark{9}}
\author{G.~Barthe\altaffilmark{9}}
\author{P.~Bendjoya\altaffilmark{10,9}}
\author{R.~Behrend\altaffilmark{14,9}}
\author{I.~Boisse\altaffilmark{2}}
\author{F.~Bouchy\altaffilmark{2,14}}
\author{H.~Boussier\altaffilmark{15,9}}
\author{M.~Bretton\altaffilmark{16,9}}
\author{D.~J.~A.~Brown\altaffilmark{7}}
\author{B.~Carry\altaffilmark{10,9}}
\author{A.~Cailleau\altaffilmark{9}}
\author{E.~Conseil\altaffilmark{17,9}}
\author{B.~Courcol\altaffilmark{2}}
\author{B.~Dauchet\altaffilmark{9}}
\author{J.-C.~Dalouzy\altaffilmark{9}}
\author{M.~Deldem\altaffilmark{18}}
\author{P.~Dubreuil\altaffilmark{9}}
\author{J.-M.~Fehrenbach\altaffilmark{13,9}}
\author{S.~Ferratfiat\altaffilmark{19,9}}
\author{R.~Girelli\altaffilmark{20,21,9}}
\author{J.~Gregorio\altaffilmark{22,9}}
\author{S.~Jaecques\altaffilmark{9}}
\author{F.~Kugel\altaffilmark{23,9}}
\author{J.~Kirk\altaffilmark{7}}
\author{O.~Labrevoir\altaffilmark{24,9}}
\author{J.-C.~Lachuri\'e\altaffilmark{13,9}}
\author{K.~W.~F.~Lam\altaffilmark{7}}
\author{P.~Le~Guen\altaffilmark{9}}
\author{P.~Martinez\altaffilmark{13,9}}
\author{L.~Maurin\altaffilmark{15,9}}
\author{J.~McCormac\altaffilmark{7}}
\author{J-B.~Pioppa\altaffilmark{25,9}}
\author{U.~Quadri\altaffilmark{20,21,26,27,9}}
\author{A.~Rajpurohit\altaffilmark{28,2}}
\author{J.~Rey\altaffilmark{14}}
\author{J.-P.~Rivet\altaffilmark{10,9}}
\author{R.~Roy\altaffilmark{29,9}}
\author{N.~C.~Santos\altaffilmark{1,30}}
\author{F.~Signoret\altaffilmark{25,9}}
\author{L.~Strabla\altaffilmark{20,21,9}}
\author{O.~Suarez\altaffilmark{10,9}}
\author{D.~Toublanc\altaffilmark{31,13,9}}
\author{M.~Tsantaki\altaffilmark{1}}
\author{P.~A.~Wilson\altaffilmark{3}}
\author{M.~Bachschmidt\altaffilmark{9}}
\author{F.~Colas\altaffilmark{32,9}}
\author{O.~Gerteis\altaffilmark{9}}
\author{P.~Louis\altaffilmark{9}}
\author{J.-C.~Mario\altaffilmark{9}}
\author{C.~Marlot\altaffilmark{9}}
\author{J.~Montier\altaffilmark{9}}
\author{V.~Perroud\altaffilmark{9}}
\author{V.~Pic\altaffilmark{9}}
\author{D.~Romeuf\altaffilmark{9}}
\author{S.~Ubaud\altaffilmark{9}}
\author{D.~Verilhac\altaffilmark{9}}

\altaffiltext{1}{Instituto de Astrof\'isica e Ci\^{e}ncias do Espa\c co, Universidade do Porto, CAUP, Rua das Estrelas, 4150-762 Porto, Portugal}
\altaffiltext{2}{Aix Marseille Universit\'e, CNRS, Laboratoire d'Astrophysique de Marseille UMR 7326, 13388, Marseille, France}
\altaffiltext{3}{Institut d'Astrophysique de Paris, UMR7095 CNRS, Universit\'e Pierre \& Marie Curie, 98bis boulevard Arago, 75014 Paris, France}
\altaffiltext{4}{Observatoire de Haute-Provence, Universit\'e d'Aix-Marseille \& CNRS, 04870 Saint Michel l'Observatoire, France}
\altaffiltext{5}{Departamento de Astrofísica, Centro de Astrobiología (CSIC-INTA), ESAC campus 28692 Villanueva de la Cañada (Madrid), Spain}
\altaffiltext{6}{European Southern Observatory (ESO), Alonso de Cordova 3107, Vitacura, Casilla 19001, Santiago de Chile (Chile)}
\altaffiltext{7}{Department of Physics, University of Warwick, Gibbet Hill Road, Coventry CV4 7AL, UK}
\altaffiltext{8}{ARC, School of Mathematics \& Physics, Queen's University Belfast, University Road, Belfast BT7 1NN, UK}
\altaffiltext{9}{European Pro/Am Network of Exoplanetary Transit Observers}
\altaffiltext{10}{Laboratoire Lagrange, UMR7239, Universit\'e de Nice Sophia-Antipolis, CNRS, Observatoire de la Cote d'Azur, 06300 Nice, France}
\altaffiltext{11}{Universit\'e Grenoble Alpes, IPAG, 38000 Grenoble, France}
\altaffiltext{12}{CNRS, IPAG, 38000 Grenoble, France}
\altaffiltext{13}{Observatoire de Belesta-en-Lauragais - Assoc. Astronomie Adagio 30 Route de Revel, 31450 Varennes, France}
\altaffiltext{14}{Observatoire Astronomique de l'Universit\'e de Gen\`eve, 51 chemin des Maillettes, 1290 Versoix, Switzerland}
\altaffiltext{15}{SAF -- Observatoire, 84450 St Saturnin les Avignon, France}
\altaffiltext{16}{Observatoire des Baronnies provencales, Observatoire Astronomique, 05150, Moydans, France}
\altaffiltext{17}{Observatoire de G\'eotopia, rue des \'ecoles, 62350 Mont-Bernenchon, France}
\altaffiltext{18}{Observatoire Les Barres, 13113 Lamanon, France}
\altaffiltext{19}{Astronomes Amateurs Aixois de l'Observatoire de Vauvenargues, 1185 chemin du Puits d'Auzon, 13126 Vauvenargues, France}
\altaffiltext{20}{Bassano Bresciano Observatory, Via San Michele,4 25020 Bassano Bresciano, Italy}
\altaffiltext{21}{Italian Supernovae Search Project, Italy}
\altaffiltext{22}{Atalaia Group -- Crow Observatory, Portalegre, Portugal}
\altaffiltext{23}{Observatoire de Dauban, 04150 Banon, France}
\altaffiltext{24}{Centre d'Astronomie, Plateau du Moulin \'a Vent, 04870, St-Michel-l'Observatoire, France}
\altaffiltext{25}{Groupement Astronomique Populaire de la R\'egion d'Antibes (GAPRA), 2 Rue Marcel-Paul 06160 Juan-Les-Pins, France}
\altaffiltext{26}{American Association of Variable Star Observers, 49 Bay State Rd., Cambridge, MA 02138, USA}
\altaffiltext{27}{Unione Astrofili Italiani -- Sezione Stelle Variabili, Italy}
\altaffiltext{28}{Astronomy and Astrophysics Division, Physical Research Laboratory, Ahmedabad 380009, India}
\altaffiltext{29}{Observatoire de Blauvac, 293 chemin de St Guillaume, 84570 Blauvac, France}
\altaffiltext{30}{Departamento de F\'isica e Astronomia, Faculdade de Ci\^encias, Universidade do Porto, Rua Campo Alegre, 4169-007 Porto, Portugal}
\altaffiltext{31}{Universit\'e de Toulouse, UPS-CNRS, IRAP, 9 Av. colonel Roche, 31028 Toulouse cedex 4, France}
\altaffiltext{32}{Institut de M\'ecanique C\'eleste et Calcul d'Eph\'em\'erides -- CNRS UMR8028, 77 Av. Denfert-Rochereau, 75014 Paris, France}

\begin{abstract}
In the present paper we report the discovery of a new hot Jupiter, EPIC211089792~b, first detected by the Super-WASP observatory and then by the \textit{K2} space mission during its campaign 4. The planet has a period of 3.25d, a mass of 0.73$\pm$0.04 \Mjup, and a radius of 1.19$\pm$0.02\Rjup. The host star is a relatively bright (V=12.5) G7 dwarf with a nearby K5V companion. Based on stellar rotation and the abundance of Lithium, we find that the system might be as young as about 450 Myr. The observation of the Rossiter-McLaughlin effect shows the planet is aligned with respect to the stellar spin. Given the deep transit (20mmag), the magnitude of the star and the presence of a nearby stellar companion, the planet is a good target for both space- and ground-based transmission spectroscopy, in particular in the near-infrared where the both stars are relatively bright.
\end{abstract}



\section{Introduction}

While small exoplanets are nowadays the most searched for objects, giant planets are still interesting to characterize for two main reasons: (1) there are still open questions that are not fully understood, such as their formation or their inflation \citep[see][and references therein]{2015arXiv151100643S} and (2) they are still the best targets for atmosphere characterization from space \citep[e.g.][]{2014Sci...346..838S} or from the ground \citep[e.g.][]{2015ApJ...802...28C}. The latter of which requires planet hosts much brighter than the typical stars targeted by the \textit{CoRoT} \citep{2006cosp...36.3749B} and \textit{Kepler} \citep{2009Sci...325..709B} space missions (V magnitude mostly between 14 and 16). 

After the failure of two of the reaction wheels, the resurrected \textit{Kepler} mission, \textit{K2} \citep{2014PASP..126..398H} is now targeting different fields of view along the ecliptic plane. \textit{K2} targets are proposed by the community through international calls. As a result, \textit{K2} is observing much brighter stars than during the prime mission (V magnitude up to 12) to allow for spectroscopic follow-up and atmosphere characterization, as well as many more M~dwarfs \citep[e.g.][]{2015ApJ...804...10C, 2015A&A...581L...7A}.

In this paper we present the discovery of a new giant planet, EPIC2011089792~b, transiting a relatively bright (V=12.5) and young ($\sim$450Myr) star in a visual binary observed during the \textit{K2} campaign 4. In Section \ref{obs}, we present the target star and the observations performed that we analyzed in Section \ref{model}. We draw our conclusion and discuss the interest of this new planet in the context of ground-based atmospheric characterization in Section \ref{conclu}.

\section{Observations and data reduction}
\label{obs}

\subsection{Ground- and space-based photometry}
\subsubsection{\textit{K2} data}

\begin{figure}[h]
\begin{center}
\includegraphics[width=\figw]{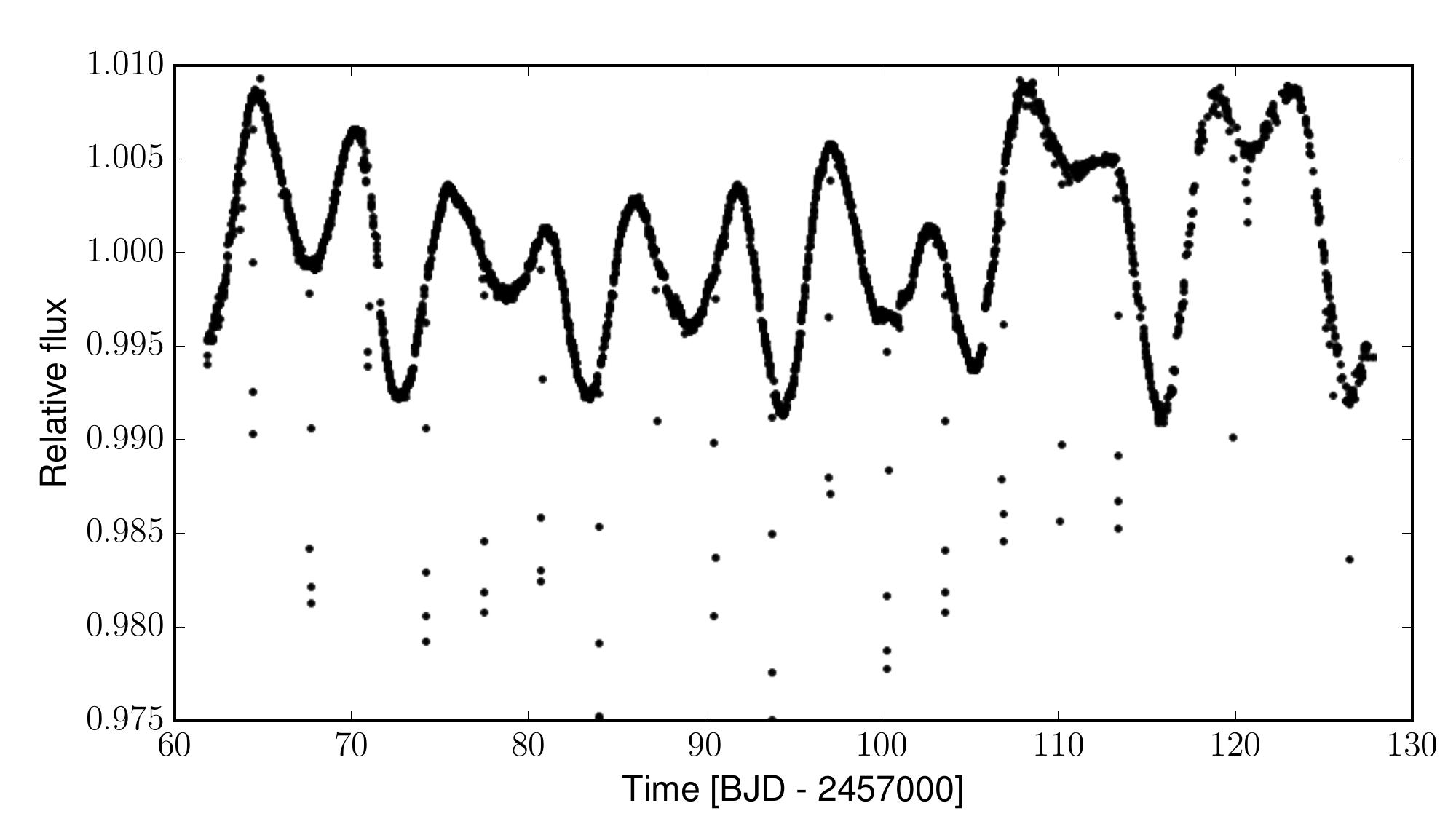}
\caption{Extracted and detrended \textit{K2} light curve of EPIC211089792.}
\label{rawLC}
\end{center}
\end{figure}

The target star EPIC2011089792\footnote{Guest Observer programme GO4007.} was observed by the \textit{Kepler} telescope from 2015-02-07 to 2015-04-23. Basic information about this target is provided in Table \ref{IDs}. We reduced the \textit{K2} raw pixel data using both the Warwick \citep{2015A&A...579A..19A, 2015A&A...582A..33A} and the LAM-K2 \citep{2015MNRAS.454.4267B} pipelines which gave similar results, except that the Warwick light curve has more noise. We therefore adopted the light curve produced by the LAM-K2 pipeline. A transiting candidate was easily detected by both pipelines as it presents a 2\%-deep transit-like event with a periodicity of about 3.258 days. This period is close to 153.5 times the integration time of the long-cadence mode of \textit{Kepler}. This means that the orbital phases covered by \textit{K2} coincide every two periods. As a consequence the transit is poorly sampled by the \textit{K2} data. We then normalized the transits by fitting a parabola to 5 hours of out-of-transit data each side of the transit. This reduced light curve is the one used for the analysis described in section \ref{model}. The star also exhibits a clear variability at the level of 1\% with a rotation period of about 11 days (see figure \ref{rawLC}). Following the work of \citet{2013ApJ...775L..11M, 2014ApJS..211...24M}, we computed the autocorrelation function of the light curve. We find the host star has a rotation period of 10.79 $\pm$ 0.02 days, which is close to three times the orbit of the planet.

\begin{table}[h]
\caption{Various identification (IDs), magnitudes, and coordinates of the target star.}
\begin{center}
\begin{tabular}{ccc}
\hline
\hline
 & value & reference\\
\hline
EPIC ID & 211089792 & \citet{2015arXiv151202643H}\\
TYC ID & 1818-1428-1 & \citet{2000AA...355L..27H}\\
RA & 04:10:40.955 & \citet{2015arXiv151202643H}\\
DEC & +24:24:07.35 & \citet{2015arXiv151202643H}\\
pmRA [mas/yr] & 4.99 & \citet{2011MNRAS.416..403F}\\
pmDEC [mas/yr] & -39.73 & \citet{2011MNRAS.416..403F}\\
\hline
\textit{Kepler} K$_{p}$ & 12.91 & \citet{2015arXiv151202643H} \\
Johnson B & 13.597 $\pm$ 0.062 & this work\\
Johnson V & 12.526 $\pm$ 0.044 & this work\\
2MASS J & 10.622 $\pm$ 0.035 & \citet{2013yCat.2328....0C}\\
2MASS H & 10.168 $\pm$ 0.041 & \citet{2013yCat.2328....0C}\\
2MASS Ks & 10.062 $\pm$ 0.034 & \citet{2013yCat.2328....0C}\\
WISE 3.4$\mu$m & 10.095$\pm$ 0.037 & \citet{2013yCat.2328....0C}\\
WISE 4.6$\mu$m & 10.142$\pm$ 0.037 & \citet{2013yCat.2328....0C}\\
WISE 12$\mu$m & 9.991 $\pm$ 0.082 & \citet{2013yCat.2328....0C}\\
\hline
\hline
\end{tabular}
\end{center}
\label{IDs}
\end{table}%

\subsubsection{Archival Super-WASP data}

A 2\% transit depth on a star of magnitude 12.5 could be easily detected from the ground. We checked the Super-WASP \citep{2006PASP..118.1407P} public data available at the NASA exoplanet archive\footnote{\url{http://exoplanetarchive.ipac.caltech.edu}} and found that the \textit{K2} star EPIC211089792 was observed at least from 2004-07-29 to 2008-03-16. This candidate was found in the WASP data with the same period, but a quick analysis of the spectrum misclassified it as an evolved star, and the star was no longer considered for precise radial velocities follow-up (Super-WASP team, private comm.). To allow a comparison of the WASP data with the \textit{K2} ones, we converted the heliocentric julian dates (HJD) from the WASP data to barycentric julian dates (BJD) in the barycentric dynamical time (TDB) reference using the online tool kindly provided by \citet{2010PASP..122..935E}. We then normalized the WASP transits by fitting a parabola in the out-of-transit parts, as for the \textit{K2} ones.

\subsubsection{Professional and amateur ground-based photometry}

To improve the sampling of the transit and the precision of the ephemeris we performed a photometric campaign to observe the transit that occurred on 2016-01-15 using a network of professional and amateur facilities in Europe (France, Portugal, and Italy). In total 19 observatories detected the same transit and are listed in Table \ref{phot}. The data were extracted using aperture photometry with the softwares \texttt{Munipack} \citep{2014ascl.soft02006H} or \texttt{AstroImageJ} \citep{2016arXiv160102622C}. We converted all the times in BJD TDB and normalised the transits as for the \textit{K2} and Super-WASP data.

\subsection{High-resolution imaging}
\label{AO}

\subsubsection{FTN seeing-limited imaging}

The SDSS DR9 \citep{2012ApJS..203...21A} images of the star EPIC211089792 revealed the presence of a close companion. To characterize this companion further, we obtained seeing-limited images with the Faulkes Telescope North (FTN), operated by the LCOGT network \citep{2013PASP..125.1031B}. We collected three exposures of 20s each in the B, V and R bands. We clearly detected a stellar companion located at about 4.3\arcsec\ and $\approx$35\degr\ North-to-East (see Fig. \ref{imaging}).

We search for archival data from the Digitalized Sky Survey. Images taken in 1949 and 1993 do not resolve this stellar companion but one can clearly see an elongated PSF. The angular separation is about 3\arcsec\ and 4.5\arcsec, and angles of $\approx$ 35\degr\, and 37\degr\ (North-to-East) in the 1949 and 1993 images, respectively. This suggests that star B is a physical companion of star A. In the rest of the paper, we refer to component A and B as the brightest and faintest stars in the system, respectively. Using aperture photometry, we find that the component B is fainter than the star A by 2.40 $\pm$ 0.03 mag, 2.14 $\pm$ 0.02 mag, and 1.78 $\pm$ 0.01 mag in the B, V, and R bands, respectively. We then used the contaminated magnitudes from the APASS catalog \citep{2014AJ....148...81M} to derive the B and V magnitudes of star A which are reported in Table \ref{IDs}.

\subsubsection{AstraLux lucky-imaging observations}

\begin{figure}[h!]
\begin{center}
\includegraphics[width=\figw]{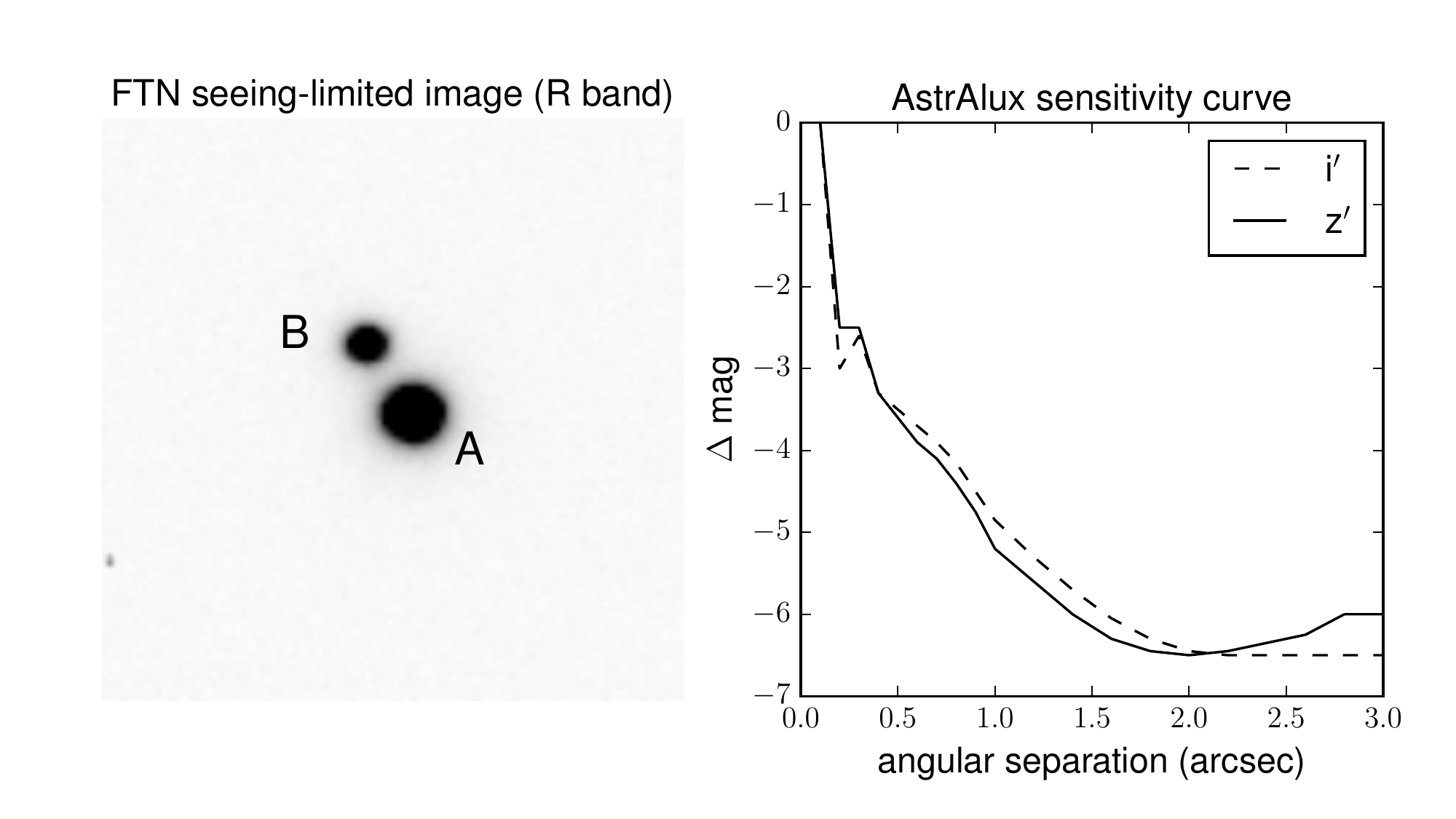}
\caption{High-resolution imaging of EPIC211089792. Left panel: FTN R-band image (30\arcsec$\times$30\arcsec, North is up, East is right). Right panel: 5-$\sigma$ sensitivity curve from AstraLux lucky-imaging observations.}
\label{imaging}
\end{center}
\end{figure}

To search for other stellar companions in the system we performed high-resolution imaging with the AstraLux lucky-imaging instrument \citep{2008SPIE.7014E..48H} mounted at the 2.2-m telescope in the Calar Alto observatory (Spain). We observed this target in the i$^{\prime}$ and z$^{\prime}$ bands on 2015-11-20 under relatively good weather conditions (seeing of around 0.9\arcsec and fair transparency with about 0.2 mag of extinction at the zenith). We obtained 90,000 frames of 30ms in full frame mode ($24\times24$\arcsec) for the i$^{\prime}$ band and 57,000 frames of 30ms for the z$^{\prime}$ band. The frames were reduced using the observatory pipeline described in \cite{2008SPIE.7014E..48H}. The pipeline performs a basic reduction of the individual frames (bias and flat-field correction), sorts them by image quality in terms of the Strehl ratio \citep{1902AN....158...89S}, then aligns and combines the best 10\% of the frames to produce the final near-diffraction limited image. We found no extra star besides star B. We used an image from the M15 globular cluster to obtain the astrometric calibration \citep[see][for details]{2014A&A...566A.103L}. In the z$^{\prime}$ band image the star B is located at $4.307\pm0.018$\arcsec. According to our aperture photometry, the star B is fainter than the star A by $1.59 \pm 0.01$ mag and $1.42 \pm 0.01$ mag in the i$^{\prime}$ and z$^{\prime}$ bands, respectively. The 5-$\sigma$ sensitivity curve of the two images within the first 3\arcsec were then computed by following the prescriptions in \citet{2014A&A...566A.103L}. The result is presented in Figure \ref{imaging}. No additional objects are found within these limits.

\subsection{Spectroscopic follow-up}

\subsubsection{CAFE}

We obtained three observations with the CAFE spectrograph \citep{2013A&A...552A..31A} mounted at the 2.2m telescope at the Calar Alto observatory (Spain). CAFE is a high-resolution spectrograph (R=63,000) with no movable pieces and a fixed wavelength coverage in the range 4000 -- 9500 \AA. The ambient conditions of the chamber where the instrument is located are continuously monitored to check for possible changes during the observations. The three spectra were reduced with the observatory pipeline, using the closest ThAr frame to perform the wavelength calibration and master bias and flats for the basic reduction. The radial velocity was extracted by using the cross-correlation technique, using a solar spectrum mask with more than two thousand  specifically selected lines \citep[see][for details]{2015A&A...577A.105L}. The radial velocities, bisector and full width half maximum (FWHM) are provided in the Table \ref{RVdata} together with their uncertainties.

\subsubsection{SOPHIE}

We observed the target star 27 times with the SOPHIE spectrograph\footnote{Program IDs: 15B.PNP.HEB} \citep{2013A&A...549A..49B} mounted at the 1.93m telescope at the Haute-Provence Observatory (France). These observations were carried out from 2015-10-08 to 2016-01-16 as part of a large programme to characterize \textit{Kepler} and \textit{K2} candidates. SOPHIE is a fiber-fed high-resolution echelle spectrographe stabilized in temperature and pressure. We used the high-efficiency (R$\sim$40,000) mode which allows about 10\ms\ precision in exposure times of less than one hour for this star. We reduced the data using the online pipeline which compute the weighted cross-correlation function (CCF) between the spectra and a numeric mask which corresponds to a G2V star \citep{1996A&AS..119..373B, 2002A&A...388..632P}. The choice of this mask is driven by the spectral type of the host star (see section \ref{model}).

We corrected the data for the charge transfer inefficiency of the CCD \citep{2009EAS....37..247B} following the procedure described in \citet{2012A&A...545A..76S}. We also correct the radial velocities from second-order instrumental drifts (not corrected by the wavelength calibration) using the radial velocities from the constant star HD56124 observed during the same nights, as done in \citet{2014A&A...571A..37S}. We list in Table \ref{RVdata} the radial velocities, bisector, and FWHM of the star with their uncertainties estimated following the methods of \citet{2010A&A...523A..88B} and \citet{2015MNRAS.451.2337S}. 

Among the 27 observations done with SOPHIE, 16 spectra were collected during the transit night of 2016-01-15 in order to detect the Rossiter-McLaughlin effect. 

The fiber of the SOPHIE spectrograph has an aperture on the sky of 3\arcsec. Depending on the seeing condition and telescope tracking precision, the light from the component B might have affected the data. If both components are physically bound, it is expected that they have nearly the same radial velocity and thus, would be unresolved spectroscopically. Using the formalism developed by \citet{2015MNRAS.451.2337S}, we estimated that in the worst case, i.e. where the star B fully contributes to the observed spectrum and that both stars have exactly the same systemic radial velocity, and given their difference of magnitudes, the observed radial velocities would be diluted by up to 2\%. This is substantially below the radial velocity photon noise we have on individual measurement and concluded that the star B should not affect significantly the radial velocities of the star A.

\subsubsection{HARPS-N}

We observed EPIC211089792A with HARPS-N\footnote{Program ID: OPT15B\_23}, a fiber-fed high-resolution (R$\sim$110,000) echelle spectrograph \citep{2012SPIE.8446E..1VC} mounted on 3.6m TNG at the La Palma Observatory (Spain). We obtained 22 spectra from 2016-01-04 to 2016-01-07 among which 19 were collected during the transit night of 2016-01-06 in order to detect the Rossiter-McLaughlin effect. As for SOPHIE, the spectra were reduced using the online pipeline and the radial velocities, bisector, and FWHM were measured on the CCF computed with a G2V mask. The fiber of HARPS-N has an aperture on the sky of 1\arcsec\ and only the star A was observed. The derived radial velocities, bisector and FWHM are listed in Table \ref{RVdata}, together with their uncertainties.

\section{Modelling of the exo-planetary system}
\label{model}

\subsection{Spectral characterization of stars A and B}

The spectral analysis was performed on the co-added HARPS-N spectra of star A. The spectroscopic parameters were derived with the ARES+MOOG method \citep[see][for details]{2014arXiv1407.5817S} which is based on the measurement on equivalent widths of iron lines with ARES \citep{2015A&A...577A..67S}. This method has been used to derive homogeneous parameters for planet-host stars \citep[e.g.]{2013A&A...556A.150S}. We corrected the \logg\ using the asteroseismic calibration of \citet{2014A&A...572A..95M}. We find that star A has a \teff\, of 5363 $\pm$ 43 K, a \logg\ of 4.49 $\pm$ 0.20 g.cm$^{-2}$, a micro-turbulence velocity \vmacro\ of 1.05 $\pm$ 0.08 \kms, and an Iron abundance \met\ of 0.16 $\pm$ 0.03 dex. Using the method described in \citet{2010A&A...523A..88B} on the CCF, we find a \vsini of 4 $\pm$ 1 \kms. We find evidence of Lithum in the co-added spectrum with an abundance of A(Li) = 1.05 $\pm$ 0.2 dex.

We attempted to take a spectrum of star B with HARPS-N but the automatic guiding of the telescope was moving to star A. Therefore, to characterize the star B, we used the magnitude differences measured by the FTN and AstraLux facilities (see section \ref{obs}). We modelled the spectral energy distribution of both stars using the BT-SETTL atmosphere models \citep{2014IAUS..299..271A} that we integrated in the B, V, R, i$^{\prime}$, and z$^{\prime}$ bands. We used the result from the spectral analysis to estimate the magnitudes of the star A and we derived the spectral parameters of the star B by fitting the observed differences of magnitude through a Markov Chain Monte Carlo algorithm (MCMC). We assumed that both stars are at the same distance, hence have the same interstellar extinction, and have the same Iron abundance. At each step of the MCMC we checked that the stellar parameters did not correspond to unphysical stars or stars older than the universe, according to the Dartmouth evolution tracks of \citet{2008ApJS..178...89D}. We find that star B has a \teff\ of 4400 $\pm$ 66 K and a \logg\ of 4.60 $\pm$ 0.04 g.cm$^{-2}$. This corresponds to a spectral type of K5V according to \citet{2000asqu.book.....C}.

This allows us to determine precisely the contamination of the star B in the light curves of star A. The contaminant fully contributes to the flux measured either by \textit{K2}, WASP, or the other professional and amateur facilities. By integrating the SED models in the \textit{Kepler}, r$^{\prime}$, and V bands, we find that the contamination is of 15.3$\pm$0.4\%, 15.4$\pm$0.4\%, and 12.8$\pm$0.4\%, respectively.

\subsection{Combined analysis of the system}

\begin{figure}[]
\begin{center}
\includegraphics[width=\figw]{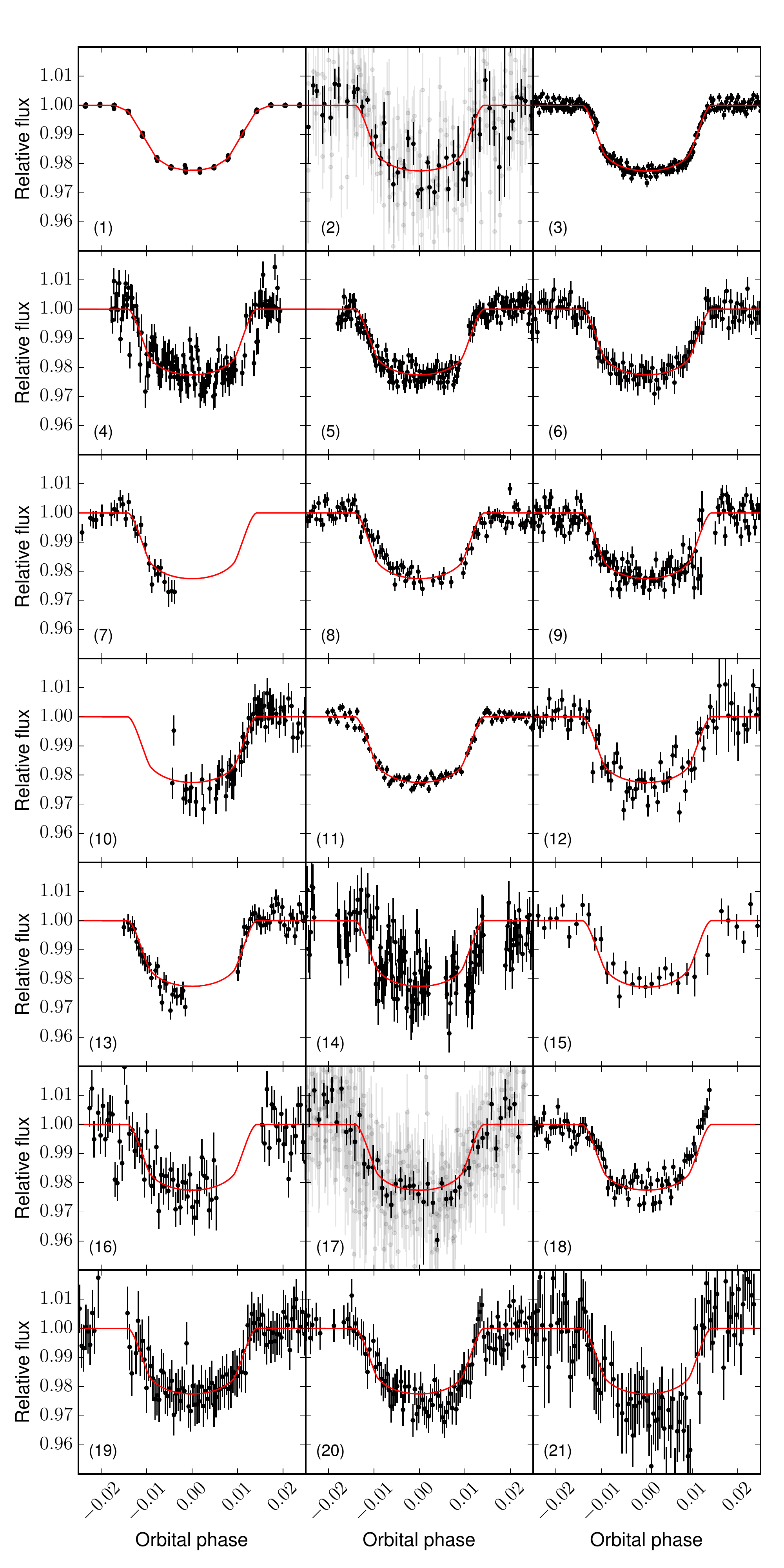}
\caption{Phase-folded transit light curves of the transit planet EPIC211089792~b. Panel 1 is the \textit{K2} data, panel 2 display the Super-WASP data and panels 3 to 21 are from the ground-based facilities listed in Table \ref{phot}. The red line is the best model found in the MCMC analysis. For panels 2 and 17, the grey dots are the raw data and the black dots are the same data binned to 0.001 in phase.}
\label{transits}
\end{center}
\end{figure}

We analyzed all the light curves, radial velocities\footnote{we excluded for this analysis the two transit nights observed by SOPHIE and HARPS-N. By doing this, we avoid biasing the system parameters with possible instrumental systematics. The analysis of the Rossiter-McLaughlin is left for the next section.} and the magnitudes (listed in the Table \ref{IDs}) of the target star using the \texttt{PASTIS} software \citep{2014MNRAS.441..983D, 2015MNRAS.451.2337S}. It models the transit light curves using a modified version of the \texttt{JKTEBOP} code \citep[and references therein]{2011MNRAS.417.2166S} that we oversampled by a factor of 10 to compensate the long integration time of the \textit{Kepler} data \citep{2010MNRAS.408.1758K}. Radial velocities are modelled with a Keplerian orbit and the SED is modelled with the BT-SETTL library \citep{2014IAUS..299..271A}. Stellar parameters are determined with the Dartmouth stellar evolution tracks and limb darkening coefficients are taken from the theoretical values of \citet{2011A&A...529A..75C}.

The statistical analysis of the data was performed with a MCMC algorithm which is fully described in \citet{2014MNRAS.441..983D}. The model is described by six free parameters for the star (\teff, \logg, \met, the systemic radial velocity $\upsilon_{0}$, the distance $d$, and the interstellar extinction $E(B-V)$), seven free parameters for the transiting planet (period $P$, epoch of first transit $T_{0}$, radial velocity amplitude $K$, the radius ratio $k_{r}$, the orbital eccentricity $e$, inclination $i$, and the angle of periastron $\omega$). We added to the model an extra source of white noise (jitter), an out-of-transit flux, and the contamination level for each of the 21 light curves listed in Table \ref{phot} which are let free in the analysis. Finally, we also added a jitter term for each of the radial velocity instruments, a radial-velocity offset between them, and a jitter term for the SED. In total, the model is composed of 82 free parameters.

We choose uninformative priors as much as possible, except for the stellar parameters that we constrained using on the results of the spectral analysis and the orbital ephemeris to speed up the convergence. We choose a Beta distribution as prior for the orbital eccentricity \citep{2013MNRAS.434L..51K}.The exhaustive list of free parameters and their prior is provided in Table \ref{PASTISparams}.

We ran 5 exploratory MCMC chains of 10$^{5}$ iterations with an initial guess randomly drawn from the joint prior distribution. We then ran 20 MCMC chains of $3\times10^{5}$ iterations started from the best solution found in the exploratory MCMC, to further explore the posterior distribution in the vicinity of the global maximum. All chains converged towards the same solution which is assumed to be the global maximum. We then removed the burn-in phase of each chain before thinning (keep only one sample per maximum correlation length among all the parameters of each chain) and merging them to obtain more than 1000 independent samples of the posterior distribution. We finally determined the median and 68.3\% confidence interval for each of the free parameters that we report in the Table \ref{PASTISparams}. Note that the uncertainties reported in this table are only the statistical ones and do not take into account the unknown uncertainties on the models.

\begin{figure}[]
\begin{center}
\includegraphics[width=\figw]{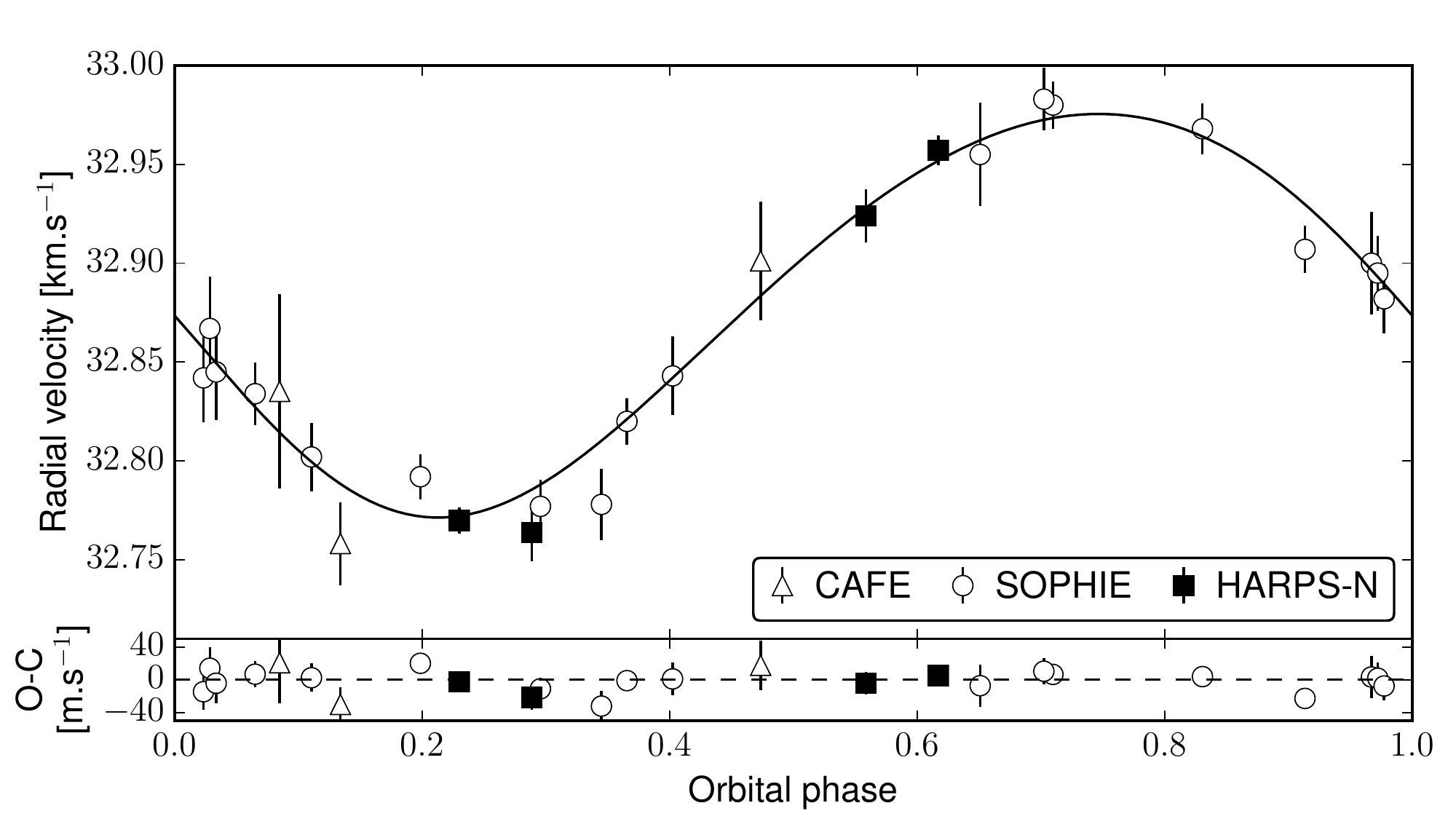}
\caption{Phase-folded radial velocities of the exoplanet EPIC211089792~b. The black line is the best model found and the bottom panel is the corresponding residuals.}
\label{RVs}
\end{center}
\end{figure}

We display in the Fig. \ref{transits} the phase-folded transit light curves from the 21 different instruments with the best-fit model. In Fig. \ref{RVs}, we plot the phase-folded radial velocity data together with the best-fit model and the residuals.

\subsection{Analysis of the Rossiter-McLaughlin effect}

We analyzed the radial velocity data obtained during the transit nights of 2016-01-06 and 2016-01-15 with HARPS-N and SOPHIE, respectively. To model the Rossiter -- McLaughlin effect, we used the formalism developed by \citet{2013A&A...550A..53B}. We are neglecting here the effects of convective blue-shift and macro-turbulence. We fit the data using the MCMC procedure as described above. We used the results of the combined analysis, listed in Table \ref{PASTISparams}, as prior for the orbital and transit parameters. We used a uniform prior for the spin-orbit angle $\lambda$ and assumed a prior for the \vsini\ which follow a normal distribution with a mean of 4\kms\ and a width of 1\kms. To account for the different integration times between HARPS-N (10 minutes) and SOPHIE (20 minutes) data, we oversampled the Rossiter--McLaughlin model to 1 minute before binning it back to the actual HARPS-N or SOPHIE cadence. This is similar to what was proposed for photometric transits by \citet{2010MNRAS.408.1758K}.

\begin{figure}[h]
\begin{center}
\includegraphics[width=\figw]{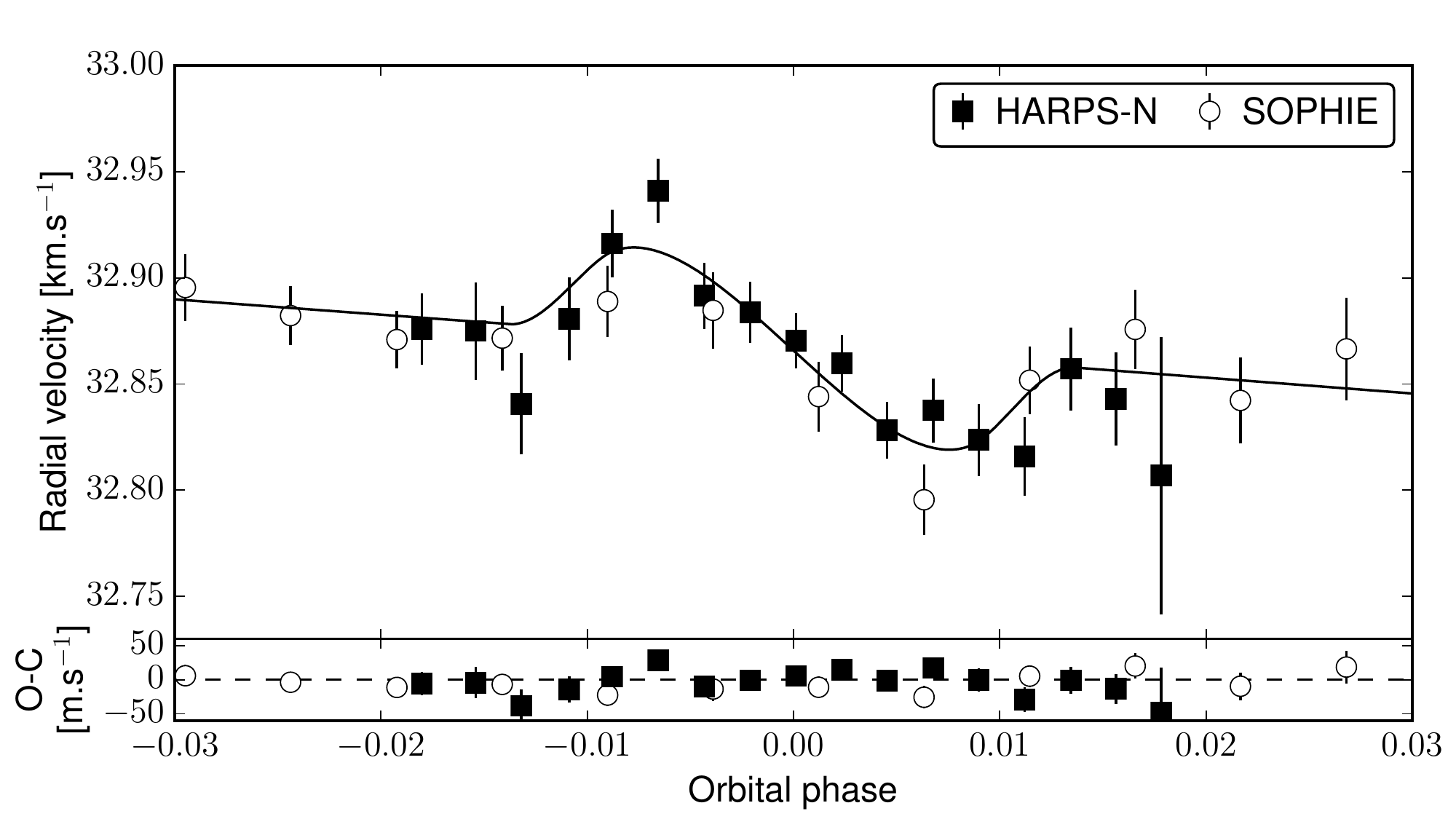}
\caption{Observation of the Rossiter--McLaughlin effect for the planet EPIC211089792~b. The black line is the best model found and the bottom panel is the corresponding residuals.}
\label{RM}
\end{center}
\end{figure}

We ran 20 chains of 3$\times$10$^{5}$ iteration each started randomly from the joint prior distribution. We analyzed the chains as previously and found that the planet is aligned relative to the stellar spin with a value of $\lambda = 1.5 \pm 8.7$\degr. The derived \vsini\ is 3.7 $\pm$ 0.5\kms. The HARPS-N and SOPHIE data are displayed in Fig. \ref{RM} together with the best-fit model. 

\subsection{Blend sanity checks}

The detection of both the reflex motion and the Rossiter -- McLaughlin effect is not enough to firmly assess the planetary nature of a candidate \citep[e.g.][Santerne et al., in prep.]{2002A&A...392..215S, 2005ApJ...619..558T}. The radial velocity variation is detected on star A, thus we can exclude star B to be the transit host. Even if it is quite unlikely, the system might still be a triple system located within the detection limits of AstraLux. According to \citet{2015MNRAS.451.2337S}, this triple system would imprint a significant variation in the bisector and/or FHWM. 

We find no variation in the bisector with $rms$ of 20\ms, 42\ms, and 15\ms\ on SOPHIE, HARPS-N, and CAFE, respectively which is compatible with the uncertainties (see Table \ref{RVdata}). The FWHM has $rms$ of 73\ms, 63\ms, and 50\ms on SOPHIE, HARPS-N, and CAFE. This is larger than the typical uncertainties (see Table \ref{RVdata}). This variability is however not correlated with the observed radial velocity variation. We concluded that this FWHM scatter is caused by the variability of the star highlighted in the \textit{K2} light curve (see Fig. \ref{rawLC}).

Finally, we reduced the SOPHIE and HARPS-N data using a binary mask corresponding to a K5 dwarf. Even if this mask does not correspond to the spectral type of the host star, it might reveal the presence of an unresolved colder stellar companion \citep{2015MNRAS.451.2337S}. The radial velocity amplitude derived with the K5V mask is consistent within the uncertainties with the one derived with a G2V mask.

From the absence of evidence of a blend in the spectroscopic and high-resolution imaging, we concluded this candidate is a \textit{bona-fide} planet.

\subsection{The planetary system}

Based on the results of the combined and Rossiter--McLaughlin analyses, we derive the physical parameters for the EPIC211089792 system and present the results in Table \ref{physics}. 

The host star has a mass of 0.94 $\pm$ 0.02 \Msun\ and a radius of 0.86 $\pm$ 0.01 \Rsun. The age derived using the Dartmouth stellar tracks \citep{2008ApJS..178...89D} indicates the system is 2.6 $\pm$ 1.2 Gyr old. The age can be however estimated as 450 $\pm$ 200 Myr based both in the A(Li) and the P$_{\rm rot}$. Both values correspond to an age older than the M34 cluster (250 Myr). The star might be younger than the Hyades cluster (or Praesepe) but few members of this 625-Myr association display either lithium abundances or rotations similar to our estimates \citep{1997AJ....114..352J, 1996A&A...310..879B, 2015A&A...583A..73B,2010A&A...515A.100J}. We can not use the activity index measured in the \ion{Ca}{2} lines as the signal-to-noise at these wavelengths is at the order of unity and thus too low for reliable measurements. A combined analysis of both stellar and planet models as done in \citet{2011A&A...527A..20G} could provide further constrain on the age of this system.

Several theoretical works showed however that the episodic accretion at the early ages can destroy Lithium \citep{2010A&A...521A..44B} as well as the accretion of planetary material to fingering convection \citep{2012ApJ...744..123T}. Considering these effects would give an even younger age for this system. Depending on the rotational evolution this star has experienced the Lithium depletion can be stronger or weaker, hence increasing the uncertainty on the age of this system.

The system is located at 185$\pm$3 pc. With a separation of $\sim$4.3\arcsec, the stellar companion (star B) has a current sky-projected separation of about 800 AU.

We find that the transiting planet has a mass of 0.73 $\pm$ 0.04 \Mjup\, and a radius of 1.19 $\pm$ 0.02 \Rjup. This gives a bulk density of 0.53 $\pm$ 0.04 g.cm$^{-3}$. EPIC211089792~b is therefore an inflated hot Jupiter. The orbit of the planet shows a 3-$\sigma$ detection of the eccentricity of 0.066 $\pm$ 0.022, but this might be caused by the effects of the stellar variability, clearly seen in the \textit{K2} light curve, affecting the radial velocity measurements. We find a sky-projected spin-orbit angle of 1.5$\pm$8.7\degr. With a stellar radius of 0.86 $\pm$ 0.01 \Rsun\ and a rotational period of 10.79 $\pm$ 0.02d, the rotational velocity of the star is 4.06 $\pm$ 0.05 \kms, which is in consistent with the \vsini\ measured with the Rossiter--McLaughlin effect of 3.7 $\pm$ 0.5 \kms. The star is therefore seen nearly edge-on and the transiting planet is well aligned with the stellar spin. This system therefore agrees with the trends reported in \citet{2012ApJ...757...18A}.

\begin{table}[h]
\caption{Physical parameters of the EPIC211089792 system}
\begin{center}
\begin{tabular}{lc}
\hline
\hline
Parameter & value and uncertainty\\
\hline
\multicolumn{2}{l}{\it Orbital parameters}\\
 &\\
Period $P$ [d] & 3.2588321 $\pm$ 1.9$\times10^{-6}$\\
Transit epoch T$_{0}$ [BJD - 2.45$\times10^{6}$] & 3219.0095 $\pm$ 2.2$\times10^{-3}$\\
Orbital eccentricity $e$ & 0.066 $\pm$ 0.022\\
Argument of periastron $\omega$ [\degr] & 132 $\pm$ 21\\
Inclination $i$ [\degr] & 86.656$^{_{+0.11}}_{^{-0.08}}$\\
Semi-major axis $a$ [AU] & 0.04217 $\pm$ 2.4$\times10^{-4}$\\
Spin-orbit angle $\lambda$ [\degr] & 1.5 $\pm$ 8.7\\
 &\\
\hline
\multicolumn{2}{l}{\it Transit \& Keplerian parameters}\\
 &\\
 System scale $a/R_{\star}$ & 10.51 $\pm$ 0.15\\
Impact parameter $b_{\rm prim}$ & 0.58 $\pm$ 0.02\\
 Transit duration T$_{14}$ [h] & 2.22 $\pm$ 0.01\\
 Planet-to-star radius ratio $k_{r}$ & 0.14188 $\pm$ 6.2$\times10^{-4}$\\
 RV amplitude $K$ [\ms] & 103.5 $\pm$ 5.4\\
 &\\
\hline
\multicolumn{2}{l}{\it Planet parameters}\\
 &\\
 Planet mass $M_{p}$ [\Mjup] & 0.73 $\pm$ 0.04\\
 Planet radius $R_{p}$ [\Rjup] & 1.19 $\pm$ 0.02\\
 Planet density $\rho_{p}$ [$\rho_{\jupiter}$] & 0.43 $\pm$ 0.03\\
 Planet density $\rho_{p}$ [g.cm$^{-3}$] & 0.53 $\pm$ 0.04\\
 Equilibrium temperature $T_{\rm eq}$ [K] & 1171 $\pm$ 10 \\
 &\\
\hline
\multicolumn{2}{l}{\it Stellar parameters}\\
 &\\
 Stellar mass $M_{\star}$ [\Msun] & 0.94 $\pm$ 0.02\\
 Stellar radius $R_{\star}$ [\Rsun] & 0.86 $\pm$ 0.01\\
 Stellar age\tablenotemark{a} $\tau$ [Gyr] & 2.6 $\pm$ 1.2\\
 Stellar age\tablenotemark{b} $\tau$ [Gyr] & 0.45 $\pm$ 0.25\\
 Distance $d$ [pc] & 185 $\pm$ 3\\
 Reddening E(B-V) [mag] & 0.19 $\pm$ 0.02\\
 Systemic RV $\upsilon_{0}$ [\kms] & 32.8786 $\pm$ 0.0044\\
 Effective temperature \teff\ [K] & 5358 $\pm$ 38\\
 Surface gravity \logg\ [g.cm$^{-2}$] & 4.540 $\pm$ 0.012\\
 Iron abundance \met\ [dex] & 0.16 $\pm$ 0.03\\
 Rotational velocity \vsini [\kms] & 3.7 $\pm$ 0.5\\
 Rotation period P$_{\rm rot}$ [d] & 10.79 $\pm$ 0.02\\
 Spectral type & G7V\\
 \hline
 \hline
\end{tabular}
\end{center}
\label{physics}
\tablecomments{All the uncertainties provided here are only the statistical ones. Errors on the models are not considered, as they are unknown.Stellar parameters are derived from the combined analysis of the data and not from the spectral analysis. We assumed 1\Rsun=695,508km, 1\Msun=1.98842$\times10^{30}$kg, 1\Rjup=71,492km, 1\Mjup=1.89852$\times10^{27}$kg, and 1AU=149,597,870.7km.}
\tablenotetext{1}{Based on the Dartmouth stellar evolution tracks.}
\tablenotetext{2}{Based on Lithium abundance and stellar rotation.}
\end{table}%

\section{Conclusion and discussion}
\label{conclu}

In this paper, we report the discovery of a new hot Jupiter co-discovered in the \textit{K2} and archival Super-WASP data. The host star is the primary of a visual binary system.

The star EPIC211089792 was observed during \textit{K2}'s campaign \#4 which also targeted both the Pleiades and the Hyades clusters. It is unlikely that this system belongs to Pleiades as estimated by \citet[membership probability of less than 2\%][]{2015A&A...577A.148B}, in agreement with \citet{2014A&A...563A..45S}. Interestingly, we note that the system has Lithium (for this stellar temperature) and Iron abundances that are compatible with the Hyades. Moreover, the systemic radial velocity of the star also agree with the Hyades \citep{1998A&A...331...81P}. However the system is too far away and the proper motion are not compatible with this cluster. We conclude it is unlikely that it belongs to the Hyades. Using the proper motion listed in Table \ref{IDs}, we find that the system has galactic velocities of U=-17\kms, V=11\kms, and W=-23\kms, and thus has a 99\% probability to belong to the thin disk.

EPIC211089792 is an inflated hot Jupiter amenable for precise spectrophotometric characterization of its atmosphere. Assuming an H$_2$ dominated atmosphere the planet gravity and equilibrium temperature would imply a scale height equal to about 418 km. The corresponding photometric precision on the transit depth measurement ($\sim\,\frac{2\,R_p\,H}{R_{\star}^2}$) is $\sim\,190$ ppm. Such a precision can be achieved  using, for example, ground-based differential spectrophotometry, given that the presence of a close-by companion will allow optimal control of systematics and subtraction of the Earth atmosphere. On a 4m class telescope, considering the effect of atmospheric scintillation and assuming an optimized observing strategy (t$\rm_{exp}\sim$15s), for one single transit event we expect to archive a precision of around half the scale height on this target. 

The two stars (target and companion) have a more favourable brightness contrast in the near-infrared (Ks$_{A}\sim$ 10.1, Ks$_{B} \sim$ 10.9). This means that EPIC211089792 should be well suited to analyze in the near infrared domain, especially from space with the \textit{JWST} \citep{2015arXiv151105528G} or \textit{ARIEL} \citep{2015DPS....4741620T}. In particular, the 1.4$\mu$m water absorption band has been found to be a powerful tracer of exoplanet atmospheric chemistry \citep{2016Natur.529...59S}. The strength of this absorption band appears to be related to the presence or absence of clouds and hazes in the atmosphere, as probed for instance by optical observations. Clear atmospheric models would imply a 1.4$\mu$m absorption depth equal to about four scale heights \citep[e.g.][]{2001ApJ...560..413H}. The equilibrium temperature of this exoplanet implies however that several compounds, in particular silicates, should be sequestered in the bottom atmosphere in the form of condensates \citep{1999ApJ...512..843B} and can potentially form cloud layers which can partially or totally mask the absorption features depending on the altitude at which they are found.

\acknowledgments

The Porto group acknowledges the support from the Funda\c{c}\~ao para a Ci\^encia e Tecnologia, FCT (Portugal) in the form of the grants, projects, and contracts UID/FIS/04434/2013 (POCI-01-0145-FEDER-007672), PTDC/CTE-AST/098528/2008, PTDC/FIS-AST/1526/2014, SFRH/BPD/76606/2011, SFRH/BPD/70574/2010, SFRH/BDP/71230/2010, IF/00169/2012, IF/00028/2014, IF/01312/2014, and POPH/FSE (EC) by FEDER funding through the program ``Programa Operacional de Factores de Competitividade - COMPETE''. AS is supported by the European Union under a Marie Curie Intra-European Fellowship for Career Development with reference FP7-PEOPLE-2013-IEF, number 627202. J.L-B acknowledges financial support from the Marie Curie Actions of the European Commission (FP7-COFUND) and the Spanish grant AYA2012- 38897-C02-01. JMA acknowledges funding from the European Research Council under the ERC Grant Agreement n. 337591-ExTrA. D.J.A. and D.P acknowledge funding from the European Union Seventh Framework programme (FP7/2007- 2013) under grant agreement No. 313014 (ETAEARTH). OD acknowledges support by CNES through contract 567133 P.A.W acknowledges the support of the French Agence Nationale de la Recherche (ANR), under program ANR-12-BS05-0012 ``Exo-Atmos''. 

This work makes use of observations from the LCOGT network. This research has made use of the VizieR catalogue access tool, CDS, Strasbourg, France. The original description of the VizieR service was published in A\&AS 143, 23. This research has made use of the NASA Exoplanet Archive, which is operated by the California Institute of Technology, under contract with the National Aeronautics and Space Administration under the Exoplanet Exploration Program. This research was made possible through the use of the AAVSO Photometric All-Sky Survey (APASS), funded by the Robert Martin Ayers Sciences Fund. This publication makes use of data products from the Wide-field Infrared Survey Explorer, which is a joint project of the University of California, Los Angeles, and the Jet Propulsion Laboratory/California Institute of Technology, funded by the National Aeronautics and Space Administration. This publication makes use of data products from the Two Micron All Sky Survey, which is a joint project of the University of Massachusetts and the Infrared Processing and Analysis Center/California Institute of Technology, funded by the National Aeronautics and Space Administration and the National Science Foundation.



{\it Facilities:} \facility{Kepler (K2)}, \facility{TNG (HARPS-N)}, \facility{OHP (SOPHIE)}, \facility{CAHA (CAFE)}. , \facility{Super-WASP}.


\begin{table*}[]
\caption{List of photometric facilities used to observe the transit on 2016-01-15. The IDs 1 and 2 are for \textit{K2} and Super-WASP, respectively.}
\begin{center}
\begin{tabular}{ccccccc}
\hline
\hline
ID & observatory / telescope & location & UAI code & aperture size & filter & observers\\
\hline
3 & C2PU & Calern observatory (FR) & 010 & 1.04m & r$^{\prime}$ & LA, JPR, PB, OS\\
4 & ADAGIO & Belesta-en-Lauragais (FR) & -- & 0.82m & R & PA, JCL, JMF, PM, DT\\
5 & TJMS & Buthiers (FR) & 199 &  0.59m & R & BD, OD, GC, JMV\\
6 & Centre Astro & St-Michel-l'Observatoire (FR) & -- & 0.58m & R & OL\\
7 & Baronnies Provencales & Moydans (FR) & B10 & 0.43m & R & MB\\
8 & TAC & Calern observatory (FR) & 010 & 0.40m & clear & FS, JBP\\
9 & Blauvac & Blauvac (FR) & -- & 0.40m & clear & RR, RB\\
10 & G\'eotopia & Mont-Bernenchon (FR) & -- & 0.32m & clear & EC\\
11 & CROW & Portalegre (PT) & -- & 0.3m & clear & JG\\
12 & Bassano Bresciano & Bassano Bresciano (IT) & 565 & 0.30m & clear & UQ, LS, RG\\
13 & AAAOV & Vauvenargues (FR) & -- & 0.3m & clear & SF \\
14 & -- & Cuq (FR) & -- & 0.3m & R & AC, VP\\
15 & Blauvac & Blauvac (FR) & -- & 0.28m & V & RR, RB\\
16 & -- & St Saturnin les Avignon (FR) & -- & 0.28m & R & HB, LM\\
17 & Dauban & Banon (FR) & -- & 0.2m & r$^{\prime}$ & FK\\
18 & Les Barres & Lamanon (FR) & K22 & 0.2m & clear & MD\\
19 & -- & Sauternes (FR) & -- & 0.2m & clear & GA\\
20 & Aspremont & Aspremont (FR) & -- & 0.2m & clear & PD, GB, SJ\\ 
21 & -- & Montebourg (FR) & -- & 0.11m & R & JCD\\
\hline
\hline
\end{tabular}
\end{center}
\label{phot}
\end{table*}%

\begin{table*}[h]
\caption{Radial velocity data for the target star EPIC211089792 with the main spectroscopic diagnoses.}
\begin{center}
\begin{tabular}{cccccccl}
\hline
\hline
Epoch & RV & $\sigma_{\rm RV}$ & BIS & $\sigma_{\rm BIS}$ & FWHM & $\sigma_{\rm FWHM}$ & Instrument\\
BJD$_{\rm TDB}$ & [\kms] & [\kms] & [\kms] & [\kms] & [\kms] & [\kms] & --\\
\hline
2457312.54420 & 32.684 & 0.021  &  -0.045  &  0.042  &  9.527  &  0.052 &   CAFE\\
2457313.65075 & 32.827 & 0.030  &  -0.013  &  0.060  &  9.455  &  0.075 &   CAFE\\
2457351.49132 & 32.761 & 0.049  &  -0.045  &  0.098  &  9.411  &  0.122 &   CAFE\\
\hline
2457304.63868  &  32.980  &  0.006  &  -0.044  &  0.012  &  10.875  &  0.016  & SOPHIE \\
2457332.61873  &  32.777  &  0.009  &  0.013  &  0.017  &  10.935  &  0.024  & SOPHIE \\
2457364.45580  &  32.834  &  0.012  &  0.002  &  0.022  &  10.897  &  0.031  & SOPHIE \\
2457378.47029  &  32.820  &  0.006  &  -0.019  &  0.011  &  10.814  &  0.015  & SOPHIE \\
2457383.51446  &  32.907  &  0.006  &  -0.015  &  0.011  &  11.004  &  0.015  & SOPHIE \\
2457384.44394  &  32.792  &  0.005  &  -0.019  &  0.009  &  10.895  &  0.013  & SOPHIE \\
2457386.50338  &  32.968  &  0.008  &  -0.026  &  0.015  &  10.862  &  0.020  & SOPHIE \\
2457400.45128  &  32.802  &  0.014  &  -0.041  &  0.025  &  10.711  &  0.035  & SOPHIE \\
2457401.40215  &  32.843  &  0.017  &  -0.011  &  0.031  &  10.742  &  0.043  & SOPHIE \\
2457402.37990  &  32.983  &  0.012  &  -0.026  &  0.021  &  10.704  &  0.029  & SOPHIE \\
2457403.24248  &  32.900  &  0.024  &  -0.013  &  0.043  &  10.858  &  0.059  & SOPHIE \\
2457403.25915  &  32.895  &  0.016  &  -0.011  &  0.028  &  10.801  &  0.039  & SOPHIE \\
2457403.27577  &  32.882  &  0.014  &  -0.044  &  0.025  &  10.796  &  0.035  & SOPHIE \\
2457403.29257  &  32.871  &  0.013  &  -0.031  &  0.024  &  10.762  &  0.034  & SOPHIE \\
2457403.30921  &  32.872  &  0.015  &  -0.032  &  0.027  &  10.729  &  0.038  & SOPHIE \\
2457403.32586  &  32.889  &  0.017  &  -0.007  &  0.030  &  10.799  &  0.041  & SOPHIE \\
2457403.34251  &  32.885  &  0.018  &  0.002  &  0.032  &  10.810  &  0.045  & SOPHIE \\
2457403.35918  &  32.844  &  0.016  &  -0.054  &  0.029  &  10.759  &  0.040  & SOPHIE \\
2457403.37584  &  32.795  &  0.017  &  -0.039  &  0.030  &  10.790  &  0.042  & SOPHIE \\
2457403.39253  &  32.852  &  0.016  &  -0.021  &  0.028  &  10.727  &  0.040  & SOPHIE \\
2457403.40920  &  32.876  &  0.019  &  -0.058  &  0.033  &  10.857  &  0.046  & SOPHIE \\
2457403.42583  &  32.842  &  0.020  &  -0.058  &  0.036  &  10.797  &  0.050  & SOPHIE \\
2457403.44257  &  32.867  &  0.024  &  0.004  &  0.043  &  10.848  &  0.060  & SOPHIE \\
2457403.45919  &  32.845  &  0.022  &  -0.059  &  0.040  &  10.825  &  0.055  & SOPHIE \\
2457404.47376  &  32.778  &  0.015  &  -0.020  &  0.027  &  10.822  &  0.038  & SOPHIE \\
2457405.47099  &  32.955  &  0.024  &  -0.008  &  0.044  &  10.700  &  0.061  & SOPHIE \\
\hline
2457392.33082  &  33.0277  &  0.0056  &  -0.0150  &  0.0085  &  8.2177  &  0.0113  & HARPS-N\\
2457393.52070  &  32.9313  &  0.0165  &  -0.0032  &  0.0247  &  8.3383  &  0.0330  & HARPS-N\\
2457393.52923  &  32.9304  &  0.0227  &  -0.0420  &  0.0341  &  8.3174  &  0.0454  & HARPS-N\\
2457393.53642  &  32.8961  &  0.0236  &  0.0638  &  0.0355  &  8.3361  &  0.0473  & HARPS-N\\
2457393.54399  &  32.9362  &  0.0193  &  0.0435  &  0.0290  &  8.2869  &  0.0387  & HARPS-N\\
2457393.55077  &  32.9716  &  0.0156  &  0.0148  &  0.0234  &  8.1960  &  0.0312  & HARPS-N\\
2457393.55803  &  32.9965  &  0.0148  &  0.0355  &  0.0223  &  8.2852  &  0.0297  & HARPS-N\\
2457393.56533  &  32.9471  &  0.0153  &  0.0373  &  0.0229  &  8.3614  &  0.0305  & HARPS-N\\
2457393.57258  &  32.9393  &  0.0142  &  0.0244  &  0.0212  &  8.3294  &  0.0283  & HARPS-N\\
2457393.57980  &  32.9259  &  0.0128  &  0.0243  &  0.0192  &  8.2970  &  0.0256  & HARPS-N\\
2457393.58709  &  32.9152  &  0.0131  &  0.0258  &  0.0197  &  8.3203  &  0.0262  & HARPS-N\\
2457393.59421  &  32.8837  &  0.0131  &  0.0437  &  0.0196  &  8.3174  &  0.0262  & HARPS-N\\
2457393.60151  &  32.8931  &  0.0148  &  -0.0204  &  0.0221  &  8.2821  &  0.0295  & HARPS-N\\
2457393.60869  &  32.8792  &  0.0167  &  0.0698  &  0.0250  &  8.3003  &  0.0334  & HARPS-N\\
2457393.61594  &  32.8714  &  0.0181  &  0.0316  &  0.0272  &  8.2850  &  0.0362  & HARPS-N\\
2457393.62327  &  32.9126  &  0.0194  &  -0.0260  &  0.0290  &  8.3519  &  0.0387  & HARPS-N\\
2457393.63035  &  32.8985  &  0.0218  &  0.1226  &  0.0327  &  8.3215  &  0.0436  & HARPS-N\\
2457393.63752  &  32.8624  &  0.0652  &  0.0442  &  0.0979  &  8.3365  &  0.1305  & HARPS-N\\
2457394.32810  &  32.8404  &  0.0044  &  0.0490  &  0.0066  &  8.1585  &  0.0088  & HARPS-N\\
2457394.51874  &  32.8343  &  0.0138  &  0.0192  &  0.0207  &  8.1709  &  0.0276  & HARPS-N\\
2457395.39857  &  32.9945  &  0.0126  &  0.0322  &  0.0189  &  8.1570  &  0.0252  & HARPS-N\\
\hline
\hline
\end{tabular}
\end{center}
\label{RVdata}
\end{table*}

\begin{table*}[]
\caption{List of free parameters used in the \texttt{PASTIS} analysis of the light curves, radial velocities and SED with their associated prior and posterior distribution. }
\begin{center}
\begin{tabular}{lcc}
\hline
\hline
Parameter & Prior\tablenotemark{a} & Posterior\\
\hline
\multicolumn{3}{l}{\it Orbital parameters}\\
 &&\\
Orbital period $P$ [d] & $\mathcal{N}(3.25883;1\times10^{-5})$ & 3.2588321 $\pm$ 1.9$\times10^{-6}$\\
Epoch of first transit T$_{0}$ [BJD$_{\rm TDB}$] - 2450000 & $\mathcal{N}(3219.0128;0.001)$ & 3219.0095 $\pm$ 2.2$\times10^{-3}$\\
Orbital eccentricity $e$ & $\beta(0.867;3.03)$ & 0.066 $\pm$  0.022\\
Argument of periastron $\omega$ [\degr] & $\mathcal{U}(0;360)$ & 132 $\pm$ 21\\
Inclination $i$ [\degr] & $\mathcal{S}(70;90)$ & 86.66$^{_{+0.11}}_{^{-0.08}}$\\
 &&\\
\hline
\multicolumn{3}{l}{\it Planetary parameters}\\
 &&\\
Radial velocity amplitude $K$ [\ms] & $\mathcal{U}(0;1000)$ & 103.5 $\pm$ 5.4\\
Planet-to-star radius ratio $k_{r}$ & $\mathcal{U}(0; 0.5)$ & 0.14188 $\pm$ 6.2$\times10^{-4}$\\
 &&\\
\hline
\multicolumn{3}{l}{\it Stellar parameters}\\
 &&\\
Effective temperature \teff\ [K] & $\mathcal{N}(5363;43)$ & 5358 $\pm$ 38\\
Surface gravity \logg\ [g.cm$^{-2}$] & $\mathcal{N}(4.49;0.20)$ & 4.540 $\pm$ 0.012\\
Iron abundance \met\ [dex] & $\mathcal{N}(0.16;0.03)$ & 0.16 $\pm$ 0.03\\
Reddening E(B-V) [mag] & $\mathcal{U}(0;1)$ & 0.19 $\pm$ 0.02\\
Systemic radial velocity $\upsilon_{0}$ [\kms] & $\mathcal{U}(-100, 100)$ & 32.8786 $\pm$ 0.0044\\
Distance to Earth $d$ [pc] & $\mathcal{P}(2;10;1000)$ & 185 $\pm$ 3\\
 &&\\
\hline
\multicolumn{3}{l}{\it Instrumental parameters\tablenotemark{b}}\\
 &&\\
CAFE radial velocity jitter [\ms] & $\mathcal{U}(0;100)$ & 35$\pm$32\\
SOPHIE radial velocity jitter [\ms] & $\mathcal{U}(0;100)$ & 12 $\pm$ 4\\
HARPS-N radial velocity jitter [\ms] & $\mathcal{U}(0;100)$ & 11$^{_{+15}}_{^{-8}}$\\
CAFE -- SOPHIE radial velocity offset [\ms] & $\mathcal{U}(-1000;1000)$ & 77 $\pm$ 30\\
HARPS-N -- SOPHIE radial velocity offset [\ms] & $\mathcal{U}(-1000;1000)$ & -71 $\pm$ 10\\
SED jitter [mag] & $\mathcal{U}(0;1)$ & 0.027 $\pm$ 0.025\\
\hline
\hline
\end{tabular}
\tablenotetext{1}{$\mathcal{N}(\mu;\sigma^{2})$ is a normal distribution with mean $\mu$ and width $\sigma^{2}$, $\mathcal{U}(a;b)$ is a uniform distribution between $a$ and $b$, $\mathcal{S}(a,b)$ is a sine distribution between $a$ and $b$, $\beta(a;b)$ is a Beta distribution with parameters $a$ and $b$, and $\mathcal{P}(n;a;b)$ is a power-law distribution of exponent $n$ between $a$ and $b$.}
\tablenotetext{2}{We did not report in this table the out-of-transit flux, jitter, and contamination for each of the 21 light curves we analysed, as they are not really meaningful. We choose uninformative priors for the two first ones and a normal prior for the latter one, to correspond with the estimated contamination and its uncertainty (see section \ref{AO}). Note that for ground-based light curves, we assumed a larger prior width (enlarged by a factor of 10 compared with the estimated error), to account for possible under/over-correction of the sky background.}
\tablerefs{The choice of prior for the orbital eccentricity is described in \citet{2013MNRAS.434L..51K}.}
\end{center}
\label{PASTISparams}
\end{table*}%


\begin{thebibliography}{}
\bibitem[Aceituno et al.(2013)]{2013A&A...552A..31A} Aceituno, J., S{\'a}nchez, S.~F., Grupp, F., et al.\ 2013, \aap, 552, A31
\bibitem[Ahn et al.(2012)]{2012ApJS..203...21A} Ahn, C.~P., Alexandroff, R., Allende Prieto, C., et al.\ 2012, \apjs, 203, 21
\bibitem[Albrecht et al.(2012)]{2012ApJ...757...18A} Albrecht, S., Winn, J.~N., Johnson, J.~A., et al.\ 2012, \apj, 757, 18
\bibitem[Allard(2014)]{2014IAUS..299..271A} Allard, F.\ 2014, IAU Symposium, 299, 271
\bibitem[Almenara et al.(2015)]{2015A&A...581L...7A} Almenara, J.~M., Astudillo-Defru, N., Bonfils, X., et al.\ 2015, \aap, 581, L7
\bibitem[Armstrong et al.(2015a)]{2015A&A...579A..19A} Armstrong, D.~J., Kirk, J., Lam, K.~W.~F., et al.\ 2015a, \aap, 579, A19
\bibitem[Armstrong et al.(2015b)]{2015A&A...582A..33A} Armstrong, D.~J., Santerne, A., Veras, D., et al.\ 2015b, \aap, 582, A33
\bibitem[Baglin et al.(2006)]{2006cosp...36.3749B} Baglin, A., Auvergne, M., Boisnard, L., et al.\ 2006, 36th COSPAR Scientific Assembly, 36, 3749
\bibitem[Baraffe \& Chabrier(2010)]{2010A&A...521A..44B} Baraffe, I., \& Chabrier, G.\ 2010, \aap, 521, A44
\bibitem[Baranne et al.(1996)]{1996A&AS..119..373B} Baranne, A., Queloz, D., Mayor, M., et al.\ 1996, \aaps, 119, 373
\bibitem[Barnes et al.(2015)]{2015A&A...583A..73B} Barnes, S.~A., Weingrill, J., Granzer, T., Spada, F., \& Strassmeier, K.~G.\ 2015, \aap, 583, A73
\bibitem[Barrado y Navascues \& Stauffer(1996)]{1996A&A...310..879B} Barrado y Navascues, D., \& Stauffer, J.~R.\ 1996, \aap, 310, 879
\bibitem[Barros et al.(2015)]{2015MNRAS.454.4267B} Barros, S.~C.~C., Almenara, J.~M., Demangeon, O., et al.\ 2015, \mnras, 454, 4267
\bibitem[Boisse et al.(2010)]{2010A&A...523A..88B} Boisse, I., Eggenberger, A., Santos, N.~C., et al.\ 2010, \aap, 523, A88
\bibitem[Borucki et al.(2009)]{2009Sci...325..709B} Borucki, W.~J., Koch, D., Jenkins, J., et al.\ 2009, Science, 325, 709
\bibitem[Bouchy et al.(2009)]{2009EAS....37..247B} Bouchy, F., Isambert, J., Lovis, C., et al.\ 2009, EAS Publications Series, 37, 247
\bibitem[Bouchy et al.(2013)]{2013A&A...549A..49B} Bouchy, F., D{\'{\i}}az, R.~F., H{\'e}brard, G., et al.\ 2013, \aap, 549, A49
\bibitem[Bou{\'e} et al.(2013)]{2013A&A...550A..53B} Bou{\'e}, G., Montalto, M., Boisse, I., Oshagh, M., \& Santos, N.~C.\ 2013, \aap, 550, A53
\bibitem[Bouy et al.(2015)]{2015A&A...577A.148B} Bouy, H., Bertin, E., Sarro, L.~M., et al.\ 2015, \aap, 577, A148
\bibitem[Brown et al.(2013)]{2013PASP..125.1031B} Brown, T.~M., Baliber, N., Bianco, F.~B., et al.\ 2013, \pasp, 125, 1031
\bibitem[Burrows \& Sharp(1999)]{1999ApJ...512..843B} Burrows, A., \& Sharp, C.~M.\ 1999, \apj, 512, 843
\bibitem[Claret \& Bloemen(2011)]{2011A&A...529A..75C} Claret, A., \& Bloemen, S.\ 2011, \aap, 529, A75 
\bibitem[Collins et al.(2016)]{2016arXiv160102622C} Collins, K.~A., Kielkopf, J.~F., \& Stassun, K.~G.\ 2016, arXiv:1601.02622
\bibitem[Cosentino et al.(2012)]{2012SPIE.8446E..1VC} Cosentino, R., Lovis, C., Pepe, F., et al.\ 2012, \procspie, 8446, 84461V
\bibitem[Cox(2000)]{2000asqu.book.....C} Cox, A.~N.\ 2000, Allen's Astrophysical Quantities,
\bibitem[Croll et al.(2015)]{2015ApJ...802...28C} Croll, B., Albert, L., Jayawardhana, R., et al.\ 2015, \apj, 802, 28
\bibitem[Crossfield et al.(2015)]{2015ApJ...804...10C} Crossfield, I.~J.~M., Petigura, E., Schlieder, J.~E., et al.\ 2015, \apj, 804, 10
\bibitem[Cutri et al.(2013)]{2013yCat.2328....0C} Cutri, R.~M., \& et al.\ 2013, VizieR Online Data Catalog, 2328
\bibitem[D{\'{\i}}az et al.(2014)]{2014MNRAS.441..983D} D{\'{\i}}az, R.~F., Almenara, J.~M., Santerne, A., et al.\ 2014, \mnras, 441, 983
\bibitem[Dotter et al.(2008)]{2008ApJS..178...89D} Dotter, A., Chaboyer, B., Jevremovi{\'c}, D., et al.\ 2008, \apjs, 178, 89
\bibitem[Eastman et al.(2010)]{2010PASP..122..935E} Eastman, J., Siverd, R., \& Gaudi, B.~S.\ 2010, \pasp, 122, 935
\bibitem[Fedorov et al.(2011)]{2011MNRAS.416..403F} Fedorov, P.~N., Akhmetov, V.~S., \& Bobylev, V.~V.\ 2011, \mnras, 416, 403
\bibitem[James et al.(2010)]{2010A&A...515A.100J} James, D.~J., Barnes, S.~A., Meibom, S., et al.\ 2010, \aap, 515, A100
\bibitem[Jones et al.(1997)]{1997AJ....114..352J} Jones, B.~F., Fischer, D., Shetrone, M., \& Soderblom, D.~R.\ 1997, \aj, 114, 352
\bibitem[Greene et al.(2015)]{2015arXiv151105528G} Greene, T.~P., Line, M.~R., Montero, C., et al.\ 2015, arXiv:1511.05528
\bibitem[Guillot \& Havel(2011)]{2011A&A...527A..20G} Guillot, T., \& Havel, M.\ 2011, \aap, 527, A20
\bibitem[H{\o}g et al.(2000)]{2000AA...355L..27H} H{\o}g, E., Fabricius, C., Makarov, V.~V., et al.\ 2000, \aap, 355, L27 
\bibitem[Hormuth et al.(2008)]{2008SPIE.7014E..48H} Hormuth, F., Hippler, S., Brandner, W., Wagner, K., \& Henning, T.\ 2008, \procspie, 7014, 701448
\bibitem[Howell et al.(2014)]{2014PASP..126..398H} Howell, S.~B., Sobeck, C., Haas, M., et al.\ 2014, \pasp, 126, 398
\bibitem[Hroch(2014)]{2014ascl.soft02006H} Hroch, F.\ 2014, Astrophysics Source Code Library, ascl:1402.006
\bibitem[Hubbard et al.(2001)]{2001ApJ...560..413H} Hubbard, W.~B., Fortney, J.~J., Lunine, J.~I., et al.\ 2001, \apj, 560, 413
\bibitem[Huber et al.(2015)]{2015arXiv151202643H} Huber, D., Bryson, S.~T., Haas, M.~R., et al.\ 2015, arXiv:1512.02643
\bibitem[Kipping(2010)]{2010MNRAS.408.1758K} Kipping, D.~M.\ 2010, \mnras, 408, 1758
\bibitem[Kipping(2013)]{2013MNRAS.434L..51K} Kipping, D.~M.\ 2013, \mnras, 434, L51
\bibitem[Lillo-Box et al.(2014)]{2014A&A...566A.103L} Lillo-Box, J., Barrado, D., \& Bouy, H.\ 2014, \aap, 566, A103
\bibitem[Lillo-Box et al.(2015)]{2015A&A...577A.105L} Lillo-Box, J., Barrado, D., Santos, N.~C., et al.\ 2015, \aap, 577, A105
\bibitem[McQuillan et al.(2013)]{2013ApJ...775L..11M} McQuillan, A., Mazeh, T., \& Aigrain, S.\ 2013, \apjl, 775, L11
\bibitem[McQuillan et al.(2014)]{2014ApJS..211...24M} McQuillan, A., Mazeh, T., \& Aigrain, S.\ 2014, \apjs, 211, 24
\bibitem[Mortier et al.(2014)]{2014A&A...572A..95M} Mortier, A., Sousa, S.~G., Adibekyan, V.~Z., Brand{\~a}o, I.~M., \& Santos, N.~C.\ 2014, \aap, 572, A95
\bibitem[Munari et al.(2014)]{2014AJ....148...81M} Munari, U., Henden, A., Frigo, A., et al.\ 2014, \aj, 148, 81
\bibitem[Pepe et al.(2002)]{2002A&A...388..632P} Pepe, F., Mayor, M., Galland, F., et al.\ 2002, \aap, 388, 632
\bibitem[Perryman et al.(1998)]{1998A&A...331...81P} Perryman, M.~A.~C., Brown, A.~G.~A., Lebreton, Y., et al.\ 1998, \aap, 331, 81
\bibitem[Pollacco et al.(2006)]{2006PASP..118.1407P} Pollacco, D.~L., Skillen, I., Collier Cameron, A., et al.\ 2006, \pasp, 118, 1407
\bibitem[Santerne et al.(2012)]{2012A&A...545A..76S} Santerne, A., D{\'{\i}}az, R.~F., Moutou, C., et al.\ 2012, \aap, 545, A76
\bibitem[Santerne et al.(2014)]{2014A&A...571A..37S} Santerne, A., H{\'e}brard, G., Deleuil, M., et al.\ 2014, \aap, 571, A37
\bibitem[Santerne et al.(2015)]{2015MNRAS.451.2337S} Santerne, A., D{\'{\i}}az, R.~F., Almenara, J.-M., et al.\ 2015, \mnras, 451, 2337
\bibitem[Santerne et al.(2016)]{2015arXiv151100643S} Santerne, A., Moutou, C., Tsantaki, M., et al.\ 2016, \aap\ in press, arXiv:1511.00643
\bibitem[Santos et al.(2002)]{2002A&A...392..215S} Santos, N.~C., Mayor, M., Naef, D., et al.\ 2002, \aap, 392, 215
\bibitem[Santos et al.(2013)]{2013A&A...556A.150S} Santos, N.~C., Sousa, S.~G., Mortier, A., et al.\ 2013, \aap, 556, A150
\bibitem[Sarro et al.(2014)]{2014A&A...563A..45S} Sarro, L.~M., Bouy, H., Berihuete, A., et al.\ 2014, \aap, 563, A45
\bibitem[Sing et al.(2016)]{2016Natur.529...59S} Sing, D.~K., Fortney, J.~J., Nikolov, N., et al.\ 2016, \nat, 529, 59
\bibitem[Sousa et al.(2011)]{2011A&A...533A.141S} Sousa, S.~G., Santos, N.~C., Israelian, G., Mayor, M., \& Udry, S.\ 2011, \aap, 533, A141
\bibitem[Sousa(2014)]{2014arXiv1407.5817S} Sousa, S.~G.\ 2014, arXiv:1407.5817
\bibitem[Sousa et al.(2015)]{2015A&A...577A..67S} Sousa, S.~G., Santos, N.~C., Adibekyan, V., Delgado-Mena, E., \& Israelian, G.\ 2015, \aap, 577, A67
\bibitem[Southworth(2011)]{2011MNRAS.417.2166S} Southworth, J.\ 2011, \mnras, 417, 2166
\bibitem[Stevenson et al.(2014)]{2014Sci...346..838S} Stevenson, K.~B., D{\'e}sert, J.-M., Line, M.~R., et al.\ 2014, Science, 346, 838
\bibitem[Strehl(1902)]{1902AN....158...89S} Strehl, K.\ 1902, Astronomische Nachrichten, 158, 89
\bibitem[Tinetti(2015)]{2015DPS....4741620T} Tinetti, G.\ 2015, AAS/Division for Planetary Sciences Meeting Abstracts, 47, 416.20
\bibitem[Torres et al.(2005)]{2005ApJ...619..558T} Torres, G., Konacki, M., Sasselov, D.~D., \& Jha, S.\ 2005, \apj, 619, 558
\bibitem[Th{\'e}ado \& Vauclair(2012)]{2012ApJ...744..123T} Th{\'e}ado, S., \& Vauclair, S.\ 2012, \apj, 744, 123
\end{thebibliography}
\end{document}